\documentstyle[proceedings,axodraw,numreferences]{crckapb} 

\begin{opening}
\title{FULFILLMENT OF THE STRONG BOOT\-STRAP CON\-DI\-TION}
\author{A. PAPA}
\institute{Dipartimento di Fisica, Universit\`a della
Calabria, \\and INFN, Gruppo collegato di Cosenza \\
87036 Arcavacata di Rende, Cosenza, Italy} 

\end{opening}

\runningtitle{FULFILLMENT OF THE STRONG BOOT\-STRAP CON\-DI\-TION}

\begin{document}

\begin{abstract}
The self-consistency of the assumption of Reggeized form of the production 
amplitudes in multi-Regge kinematics, which are used in the derivation of the 
BFKL equation, leads to strong bootstrap conditions. 
The fulfillment of these conditions opens the way to a rigorous proof of the 
BFKL equation in the next-to-leading approximation. The strong bootstrap 
condition for the kernel of the BFKL equation for the octet color 
state of two Reggeized gluons is one of these conditions. We show that 
it is satisfied in the next-to-leading approximation.
\end{abstract}

% The \begin{document} command comes after the \end{opening}
% command.

\section{Gluon Reggeization and bootstrap conditions}

The property of gluon Reggeization plays an essential role in the derivation 
of the Balitsky-Fadin-Kuraev-Lipatov (BFKL) equation~\cite{Fad02}
for the cross sections at 
high energy $\sqrt s$ in perturbative QCD. The simplest realization of the 
gluon Reggeization is in the elastic process 
$ A + B \longrightarrow A^\prime + B^\prime $ with exchange of gluon quantum 
numbers in the $t$-channel, whose amplitude in the Regge limit 
(i.e. for $s\rightarrow\infty$ and $|t|$ not growing with $s$)
takes the form
\begin{equation}
\left({{\cal A}_8^-}\right)^{A^\prime B^\prime}_{AB} = 
\Gamma^c_{A^\prime A}\:\left[\left({-s\over -t}\right)^{j(t)}-\left({s\over 
-t}\right)^{j(t)}\right]\:\Gamma^c_{B^\prime B}\;.
\label{elast_ampl_8}
\end{equation}
Here $c$ is a color index, $\Gamma^c_{P^\prime P}$ are the 
particle-particle-Reggeon (PPR) vertices, not depending on $s$,
and $j(t)=1+\omega(t)$ is the Reggeized gluon trajectory.

In the leading logarithmic approximation (LLA), which means resummation of the 
terms of the form $\alpha_s^{n} (\ln s)^n$, this form of the amplitude 
has been proved~\cite{Fad02}. 
In the next-to-leading approximation (NLA), which means resummation of 
the terms $\alpha_s^{n+1} (\ln s)^n$, the form~(\ref{elast_ampl_8}) has been
checked in the first three orders of perturbation theory and is
only assumed to be valid to all orders. 

On the other hand, the amplitude for the elastic scattering process 
$ A + B \longrightarrow A^\prime + B^\prime $ (for any color representation
in the $t$-channel resulting from the composition of two color octet 
representations) can be determined from $s$-channel unitarity, by 
expressing its imaginary part in terms of the inelastic amplitudes
$A + B \longrightarrow \tilde A + \tilde B + \{n\}$ and 
$A^\prime + B^\prime \longrightarrow \tilde A + \tilde B + \{n\}$, and then
by reconstructing the full amplitude by use of the dispersion relations. 
It turns out (see for instance~\cite{Fad02,FF98}) that this amplitude can 
be written as
\[
({\cal A}_{\cal R})_{AB}^{A^{\prime }B^{\prime }}=
\frac{i \, s}{\left( 2\pi \right) ^{D-1}}\int \frac{d^{D-2}q_{1}}{\vec{q}
_{1}^{\:2} \vec{q}_{1}^{\:\prime\:2}}\int \frac{d^{D-2}q_{2}}{\vec{
q}_{2}^{\:2} \vec{q}_{2}^{\:\prime\: 2}}\int_{\delta-i\infty}^{\delta+i\infty} 
\frac{d\omega}{\mbox{sin}(\pi\omega)}
\left[ \left( \frac{-s}{s_{0}}\right)^{\omega}
\right.
\]
\begin{equation}
\left. -\tau\left( \frac{s}{s_{0}}\right)^{\omega}\right]
\sum_{ {\cal R},\nu} \Phi _{A^{\prime
}A}^{\left( {\cal R},\nu \right) }\left( \vec{q}_{1},\vec{q};s_{0}\right)
G_{\omega }^{\left( {\cal R}\right) }\left( \vec{q}_{1},\vec{q}_{2},\vec{q}\right)
\Phi _{B^{\prime }B}^{\left( {\cal R},\nu \right) }
\left( -\vec{q}_{2},-\vec{q};s_{0}\right),
\label{elast_ampl_R}
\end{equation}
where $D=4+2\epsilon$ is the space-time dimension, 
${\cal A}_{\cal R}$ stands for the scattering amplitude with
the representation ${\cal R}$ of the color group in the $t$-channel, 
the index $\nu$ enumerates the states in the irreducible representation 
${\cal R}$ and $G_{\omega }^{\left( {\cal R}\right)}$ is the Mellin 
transform of the Green function for the Reggeon-Reggeon scattering.
The signature $\tau$ is positive (negative) for symmetric (antisymmetric) 
representation ${\cal R}$. The parameter $s_0$ is an arbitrary energy scale 
introduced in order to define the partial wave expansion of the scattering 
amplitudes through the Mellin transform. The dependence on this parameter 
disappears in the full expressions for the amplitudes. 
$\Phi_{P^\prime P}^{\left( {\cal R},\nu \right)}$ are the so-called impact factors.

The Green function obeys the generalized BFKL equation 
\begin{eqnarray}
\omega G_{\omega }^{\left( {\cal R}\right) }\left( \vec{q}_{1},\vec{q}_{2},
\vec{q}\right) &=& \vec{q}_{1}^{\:2}\vec{q}_{1}^{\:\prime\:2}
\delta ^{\left(
D-2\right) }\left( \vec{q}_{1}-\vec{q}_2\right) \nonumber \\
&+&\int \frac{d^{D-2}{q}_r}{\vec{q}_r^{\:\prime\:2}\vec{q}_r^{\:2}}
{\cal K}^{\left( {\cal R}\right) }\left( \vec{q}_{1},\vec{q}_r;\vec{q}
\right) G_{\omega }^{\left( {\cal R}\right) }\left( \vec{q}_r,\vec{q}
_{2};\vec{q}\right)\;,
\label{BFKL}
\end{eqnarray}
where we have introduced the notation $q_i^\prime \equiv q-q_i$. 
The kernel ${\cal K}^{( {\cal R})}$ consists 
of two parts, a ``virtual'' part, related with the Reggeized gluon trajectory, and 
a ``real'' part, related to particle production:
\begin{eqnarray}
{\cal K}^{\left( {\cal R}\right) }\left( \vec{q}_{1},\vec{q}_{2};\vec{q}
\right) &=& \left[ \omega \left( -\vec{q}_{1}^{\:2}\right) +\omega \left( -
\vec{q}_{1}^{\:\prime\:2}\right) \right] \: \vec{q}_{1}^{\:2} \:
\vec{q}_{1}^{\:\prime\:2}\: \delta ^{\left( D-2\right)
}\left( \vec{q}_{1}-\vec{q}_{2}\right) \nonumber \\
&+&{\cal K}_{r}^{\left( {\cal R}\right)
}\left( \vec{q}_{1},\vec{q}_{2};\vec{q}\right)  \;.
\label{kernel}
\end{eqnarray}
In the LLA, the Reggeized gluon trajectory is needed at 1-loop accuracy and the 
only contribution to the ``real'' part of the kernel is from the production 
of one gluon at Born level in the collision of two Reggeons. 
In the NLA, the Reggeized gluon trajectory is needed at 2-loop accuracy and the 
``real'' part of the kernel takes contributions from one-gluon production at 
1-loop level and from two-gluon and $q \overline q$-pair production at Born 
level. The representation~(\ref{elast_ampl_R})
for the elastic amplitude must reproduce with NLA accuracy the representation
(\ref{elast_ampl_8}), in the case of exchange of gluon quantum numbers (i.e.
color octet representation and negative signature) in the $t$-channel.
This leads to two so-called ``soft'' bootstrap conditions~\cite{FF98}:
the first of them involves the kernel of the generalized non-forward
BFKL equation in the octet color representation;
the second one involves the impact factors in the octet color representation.
Besides providing a stringent check of the gluon Reggeization in the NLA, 
the soft bootstrap conditions are important since they test, at least in part, the 
correctness of the year-long calculations which lead to the NLA BFKL equation.
The first bootstrap condition has been verified at arbitrary space-time dimension, 
for the part concerning the quark contribution to the kernel (in massless 
QCD)~\cite{FFP99} and, in the $D \rightarrow 4$ limit, for the part concerning the
gluon contribution to the kernel~\cite{FFK00}. 
The second bootstrap condition is process-dependent and should therefore be checked
for every new impact factor which is calculated\footnote{See, however, 
Ref.~\cite{Kot02} where a general proof for arbitrary process has been 
sketched.}. It has been verified explicitly at arbitrary space-time dimension for 
quark and gluon impact factors in QCD with massive quarks~\cite{FFKP00_Q,FFKP00_G}.

The derivation of the BFKL equation in the NLA involves also the assumption of 
Reggeized form for production amplitudes in multi-Regge and quasi-multi-Regge 
kinematics. The compatibility of these amplitudes with $s$-channel unitarity 
leads to the so-called strong bootstrap conditions~\cite{Fad02}. The fulfillment of 
these conditions opens the way to a proof of the gluon Reggeization in the 
NLA~\cite{Fad02}. Among these conditions, there appear the two ones suggested by 
Braun and Vacca~\cite{BV99} and derived from the assumption of gluon Reggeization 
in the (unphysical) particle-Reggeon scattering amplitude with gluon quantum numbers
in the $t$-channel~\cite{FFKP00}. They can be presented in the form (octet color 
representation understood)
\begin{equation}
\int\frac{d^{D-2}q_2}{\vec q_2^{\:2}\vec q_2^{\:\prime \:2}}{\cal K}(\vec q_1,
\vec q_2;\vec q)R(\vec q_2,\vec q) =\omega(t)R(\vec q_1,\vec q)\;,
\label{boot1}
\end{equation}
\begin{equation}
\Phi^{a}_{A^\prime A}(\vec q_1,\vec q) =\frac{-ig\sqrt{N}}{2}\Gamma_{A^\prime
A}^{a}(q)R(\vec q_1,\vec q)~.
\label{boot2}
\end{equation}
The last condition fixes the process dependence of the impact factor: 
it is proportional to the corresponding effective vertex, with a 
universal coefficient function $R$~\cite{FFKP00}.
The II strong bootstrap condition has been verified in the case of 
gluon and quark impact factors in the NLA~\cite{FFKP00}.
Let us consider the I strong bootstrap condition. Using Eq.~({\ref{kernel}), 
it can be written in the form
\begin{equation}
\int\frac{d^{D-2}q_2}{\vec q_2^{\:2}\vec q_2^{\:\prime \:2}}{\cal K}_r(\vec q_1,
\vec q_2;\vec q)R(\vec q_2,\vec q) =\left(\omega(t)-\omega(t_1)-\omega(t_1^\prime)
\right)R(\vec q_1,\vec q)~, 
\label{boot1'}
\end{equation}
where $t = q^2=-\vec q^{\:2}$, $t_i=q_i^2=-\vec q_i^{\:2}$ and $t_i^\prime=
q_i^{\prime\:2}=-\vec q_i^{\:\prime\:2}$, with $i=1,2$.
At the leading order, we have:
\begin{equation}
\int\frac{d^{D-2}q_2}{\vec q_2^{\:2}\vec q_2^{\:\prime \:2}}{\cal K}^{(0)}_r
(\vec q_1, \vec q_2;\vec q)=\omega^{(1)}(t)-\omega^{(1)}(t_1)-\omega^{(1)}
(t_1^\prime)\;,
\label{boot1_LO}
\end{equation}
with
\[
{\cal K}_r^{(0)}(\vec q_1,\vec q_2; \vec q) = \frac{g^2 N_c}{2(2\pi)^{D-1}}\:
f_B(\vec q_1,\vec q_2; \vec q) \;,
\]
\[
f_B(\vec q_1,\vec q_2; \vec q) = \frac{\vec q_1^{\:2} \vec q_2^{\:\prime\:2}
+\vec q_2^{\:2} \vec q_1^{\:\prime\:2}}{\vec k^{\:2}}-\vec q^{\:2}\;,
\;\;\; \vec k =\vec q_1 - \vec q_2
\]
and
\[
\omega^{(1)}(t) =\frac{g^2N_c t}{2(2\pi)^{D-1}} \int\frac{d^{D-2}q_1}{\vec
q_1^{\:2}\vec q_1^{\:\prime \:2}}=-g^2 \frac{N_c \Gamma(1-\epsilon)}{(4 \pi)^{D/2}}
\frac{\Gamma^2(\epsilon)}{\Gamma(2\epsilon)} (\vec q^{\:2})^\epsilon \;.
\]
It is trivial to see that the leading order bootstrap condition~(\ref{boot1_LO})
is satisfied. In the NLA (taking into account Eq.~(\ref{boot1_LO})), we have instead
\[
\int\frac{d^{D-2}q_2}{\vec q_2^{\:2} \vec q_2^{\:\prime\:2}}
\biggl[{\cal K}_r^{(1)}(\vec q_1,\vec q_2, \vec q)+
{\cal K}_r^{(0)}(\vec q_1,\vec q_2, \vec q)\: \left(R^{(1)}(\vec q_2, \vec q)-
 R^{(1)}(\vec q_1, \vec q)\right)
\]
\begin{equation}
= \omega^{(2)}(t) -\omega^{(2)}(t_1)-\omega^{(2)}(t_1^\prime)~.
\label{boot1''}
\end{equation}
The fulfillment of the above relation for the quark part has been already 
verified~\cite{BV99,FFKP00}, while the fulfillment for the gluon part is much more 
complicated and is the subject of this paper. 

\section{Proof of the I strong bootstrap condition for the gluon part of the kernel}

The ingredients for the calculation are the following (gluon part is understood 
everywhere):
\begin{eqnarray}
\omega^{(2)}(t) &=&\left[\frac{g^2N_c \Gamma(1-\epsilon)(\vec q^{\:2})^\epsilon}
{(4 \pi)^{D/2}\epsilon}\right]^2
\left[\frac{11}{3}+\left(2 \psi'(1)-\frac{67}{9}\right)\epsilon \right.
\nonumber \\
&+&\left.\left(\frac{404}{27}+\psi''(1)-\frac{22}{3}\psi'(1)\right)\epsilon^2\right]
+O(\epsilon)]\;,
\label{omega2}
\end{eqnarray}
\[
R^{(1)}(\vec q_1, \vec q) = \frac{\omega^{(1)}(t)}{2}
\left[\frac{\epsilon \Gamma(1+2\epsilon)(\vec q^{\:2} )^{1-\epsilon}}
{2\Gamma^2(1+\epsilon)}\int\frac{d^{D-2}k}{\Gamma(1-\epsilon)\pi^{1+\epsilon}}
\frac{\ln(\vec q^{\:2} /\vec k^{\:2})}{(\vec k - \vec q_1)^2(\vec k +
\vec q_1^{\:\prime})^2}
\right.
\]
\[
\left.
+\!\left(\!\left(\frac{\vec q_1^{\:2}}{\vec q^{\:2}}\right)^\epsilon\!+\!
\left(\frac{\vec q_1^{\:\prime\:2}}{\vec q^{\:2}}\right)^\epsilon\!-1\! \right)\!
\left(\frac{1}{2\epsilon}\!+\psi(1+2\epsilon)-\psi(1+\epsilon)+\frac{11+7\epsilon}
{2(1+2\epsilon)(3+2\epsilon)}\right) \right.
\]
\begin{equation}
\left.
-\frac{1}{2\epsilon}+\psi(1)
+\psi(1+\epsilon)-\psi(1-\epsilon)-\psi(1+2\epsilon)
\right]\;, 
\end{equation}
\[
{\cal K}_r^{(1)} = \frac{\bar g^4}{\pi^{1+\epsilon}
\Gamma(1-\epsilon)} \: \biggl( {\cal K}_1 + {\cal K}_2 + 
{\cal K}_3 \biggr)\;,
\;\;\;\;\;\;\;\;\;\;
\bar g^2 \equiv \frac{g^2 N_c \Gamma(1-\epsilon)}{(4 \pi)^{D/2}}\;,
\]
with
\[
{\cal K}_1 = -f_B(\vec q_1,\vec q_2; \vec q) \frac{(\vec k^{\:2})^\epsilon}
{\epsilon}
\left[\frac{11}{3}+\left(2 \psi'(1)-\frac{67}{9}\right)\epsilon \right.
\]
\begin{equation}
\left. 
+ \left(\frac{404}{27}+7 \psi''(1)-\frac{11}{3}\psi'(1)\right)\epsilon^2\right]\;,
\end{equation}
\[
{\cal K}_2 = 
\left\{\vec q^{\:2}\left[ \frac{11}{6}\ln\left(\frac{\vec q_1^{\:2}
\vec q_2^{\:2}}{\vec q^{\:2} \vec k^{\:2}}\right)
+\frac{1}{4}\ln\left(\frac{\vec q_1^{\:2}}{\vec q^{\:2}}\right)
\ln\left(\frac{\vec q_1^{\:\prime\:2}}{\vec q^{\:2}}\right)
+\frac{1}{4}\ln\left(\frac{\vec q_2^{\:2}}{\vec q^{\:2}}\right)
\ln\left(\frac{\vec q_2^{\:\prime\:2}}{\vec q^{\:2}}\right)\right.\right.
\]
\[
\left. +\frac{1}{4}\ln^2\left(\frac{\vec q_1^{\:2}}{\vec q_2^{\:2}}\right)\right]
-\frac{\vec q_1^{\:2} \vec q_2^{\:\prime\:2}+\vec q_2^{\:2}
\vec q_1^{\:\prime\:2}}{2 \vec k^{\:2}}\ln^2\left(\frac{\vec q_1^{\:2}}
{\vec q_2^{\:2}}\right) 
\]
\begin{equation}
\left.
+\frac{\vec q_1^{\:2} \vec q_2^{\:\prime\:2}-\vec q_2^{\:2}
\vec q_1^{\:\prime\:2}}{\vec k^{\:2}}\ln\left(\frac{\vec q_1^{\:2}}{\vec q_2^{\:2}}
\right)\left(\frac{11}{6}-\frac{1}{4}\ln\left(\frac{\vec q_1^{\:2}
\vec q_2^{\:2}}{\vec k^4}\right)\right)\right\}+\biggl\{\vec q_i\leftrightarrow
\vec q_i^{\:\prime}\biggr\}\;,
\end{equation}
\[
{\cal K}_3 = \left\{\frac{1}{2}\left[\vec q^{\:2}(\vec k^{\:2}-\vec q_1^{\:2}
-\vec q_2^{\:2})+2 \vec q_1^{\:2} \vec q_2^{\:2}
- \vec q_1^{\:2} \vec q_2^{\:\prime\:2}
- \vec q_2^{\:2} \vec q_1^{\:\prime\:2}
\right.\right.
\]
\[
\left.\left.
+\frac{\vec q_1^{\:2} \vec q_2^{\:\prime\:2}-\vec q_2^{\:2}\vec q_1^{\:\prime\:2}}
{\vec k^{\:2}}(\vec q_1^{\:2}-\vec q_2^{\:2})\right]
\int_0^1\frac{dx}{(\vec q_1(1-x)+\vec q_2 x)^2}\ln\left(\frac{\vec q_1^{\:2}
(1-x)+\vec q_2^{\:2} x}{\vec k^{\:2} x(1-x)}\right)\!\right\}
\]
\begin{equation}
+\biggl\{\vec q_i \leftrightarrow \vec q_i^{\:\prime}\biggr\}\;.
\label{K3}
\end{equation}
The integral in $R^{(1)}(\vec q_1, \vec q)$ is already known with 
$O(\epsilon^0)$ accuracy (see the Appendix of Ref.~\cite{FFKP00}):
\[
(\vec q^{\:2})^{1-\epsilon}\!\int\frac{d^{D-2}k}{\Gamma(1-\epsilon)\pi^{1+\epsilon}}
\frac{\ln(\vec q^{\:2} /\vec k^{\:2})}{(\vec k - \vec q_1)^2(\vec k
+ \vec q_1^{\:\prime})^2}
= -\frac{1}{\epsilon}\ln\!\left(\frac{\vec q_1^{\:2} \vec q_1^{\:\prime\:2}}
{(\vec q^{\:2})^2}\right)
-\frac{1}{2} \ln^2\!\!\left(\frac{\vec q_1^{\:2}}{\vec q_1^{\:\prime\:2}}\right).
\]
Using the above result, the I strong bootstrap condition~(\ref{boot1''}) takes 
the following form, with $O(\epsilon^0)$ accuracy:
\[
\frac{(\vec q^{\:2})^{-2\epsilon}}{\pi^{1+\epsilon}\Gamma(1-\epsilon)}
\int\frac{d^{D-2}q_2}{\vec q_2^{\:2} \vec q_2^{\:\prime\:2}}
\left[{\cal K}_1+{\cal K}_2+{\cal K}_3 +f_B\left(-\frac{11}{6}
\ln\left(\frac{\vec q_2^{\:2} \vec q_2^{\:\prime\:2}}{\vec q_1^{\:2}
\vec q_1^{\:\prime\:2}}\right)\right.\right.
\]
\[
\left.\left.
+\frac{1}{2}\ln\left(\frac{\vec q_1^{\:2}}
{\vec q^{\:2}}\right)\ln\left(\frac{ \vec q_1^{\:\prime\:2}}{\vec q^{\:2}}
\right)
-\frac{1}{2}\ln\left(\frac{\vec q_2^{\:2}}{\vec q^{\:2}}\right)
\ln\left(\frac{ \vec q_2^{\:\prime\:2}}{\vec q^{\:2}}\right)\right)\right]
\]
\[
=-\frac{1}{\epsilon^2}
\left[\frac{11}{3}+\left(2 \psi'(1)-\frac{67}{9}\right)
\epsilon+\left(\frac{404}{27}+\psi''(1)-\frac{22}{3}\psi'(1)\right)\epsilon^2\right]
\]
\begin{equation}
-\frac{2}{\epsilon}
\left[\frac{11}{3}+\left(2 \psi'(1)-\frac{67}{9}\right)\epsilon \right]
\ln\left(\frac{\vec q_1^{\:2} \vec q_1^{\:\prime\:2}}{(\vec q^{\:2})^2}\right)
-\frac{22}{3}\left(\ln^2\left(\frac{\vec q_1^{\:2}}{\vec q^{\:2}}\right)+
\ln^2\left(\frac{\vec q_1^{\:\prime\:2}}{\vec q^{\:2}}\right) \right)
\label{boot1'''}
\end{equation}
For details on the calculation of the remaining integrals, we refer to~\cite{FP02};
here we merely quote the final results. The integral from ${\cal K}_1$ is
\[
\frac{(\vec q^{\:2})^{-2\epsilon}}{\pi^{1+\epsilon}\Gamma(1-\epsilon)}
\int\frac{d^{D-2}q_2}{\vec q_2^{\:2} \vec q_2^{\:\prime\:2}}
\left(\frac{\vec q_1^{\:2}\vec q_2^{\:\prime\:2}+\vec q_2^{\:2}\vec
q_1^{\:\prime\:2}}{(\vec q_1-\vec q_2)^2}-\vec q^{\:2}\right)
\frac{[(\vec q_1-\vec q_2)^2]^\epsilon}{\epsilon}
\]
\[
= \frac{1}{\epsilon^2}
\left[1+2\epsilon \ln\left(\frac{\vec q_1^{\:2} \vec q_1^{\:\prime\:2}}{(\vec
q^{\:2})^2}\right)
+\epsilon^2\left(2\ln^2\left(\frac{\vec q_1^{\:2}}{\vec q^{\:2}}\right)+
2\ln^2\left(\frac{\vec q_1^{\:\prime\:2}}{\vec q^{\:2}}\right) 
\right.\right.
\]
\begin{equation}
\left.\left. +\ln\left(\frac{\vec 
q_1^{\:2}}{\vec q^{\:2}}\right)\ln\left(\frac{ \vec q_1^{\:\prime\:2}}
{\vec q^{\:2}}\right)-\psi^{\prime}(1)\right)\right] +O(\epsilon)\;.
\label{int_K1}
\end{equation}
Among the remaining integrals, there are four which can be calculated in a 
straightforward way using the generalized Feynman parametrization 
for arbitrary space-time dimension (although we quote here only the 
result with $O(\epsilon^0)$ accuracy):
\[
I_1(\vec q) =
\frac{(\vec q^{\:2})^{1-\epsilon}}{\pi^{1+\epsilon}\Gamma(1-\epsilon)}
\int d^{D-2}q_2 \;\frac{1}{\vec q_2^{\:2} (\vec q_2-\vec q)^2}
= \frac{2}{\epsilon}+O(\epsilon)\;,
\]
\[
I_2(\vec q) =
\frac{(\vec q^{\:2})^{1-\epsilon}}{\pi^{1+\epsilon}\Gamma(1-\epsilon)}
\int d^{D-2}q_2 \;\frac{\ln(\vec q_2^{\:2}/\vec q^{\:2})}
{\vec q_2^{\:2} (\vec q_2-\vec q)^2}
%\]
%\[
%= \frac{\Gamma^2(\epsilon)}{\Gamma(2\epsilon)}
%\biggl[\psi(\epsilon)-\psi(2\epsilon)+\psi(1)-\psi(1-\epsilon)\biggr]
= -\frac{1}{\epsilon^2}+\psi'(1)+O(\epsilon)\;,
\]
\[
I_3(\vec q) =
\frac{(\vec q^{\:2})^{1-\epsilon}}{\pi^{1+\epsilon}\Gamma(1-\epsilon)}
\int d^{D-2}q_2 \;\frac{\ln^2\left(\vec q_2^{\:2}/\vec q^{\:2}\right)}
{\vec q_2^{\:2} (\vec q_2-\vec q)^2}
%\]
%\[
%= \frac{\Gamma^2(\epsilon)}{\Gamma(2\epsilon)}
%\biggl[\biggl(\psi(\epsilon)-\psi(2\epsilon)+\psi(1)-\psi(1-\epsilon)\biggr)^2
%+\psi'(\epsilon)-\psi'(2\epsilon)
%\]
%\[
%- \psi'(1)+\psi'(1-\epsilon)\biggr]
= \frac{2}{\epsilon^3} - \frac{2\psi'(1)}{\epsilon}-2 \psi''(1)+O(\epsilon)\;,
\]
\[
I_4(\vec q)\!=\!
\frac{(\vec q^{\:2})^{1-\epsilon}}{\pi^{1+\epsilon}\Gamma(1-\epsilon)}
\!\!\int \!\!d^{D-2}q_2 \frac{\ln\left(\vec q_2^{\:2}/\vec q^{\:2}\right)
\ln\left((\vec q_2-\vec q)^2/\vec q^{\:2}\right)}
{\vec q_2^{\:2} (\vec q_2-\vec q)^2}
%\]
%\[
%= \frac{\Gamma^2(\epsilon)}{\Gamma(2\epsilon)}
%\biggl[\biggl(\psi(\epsilon)-\psi(2\epsilon)+\psi(1)-\psi(1-\epsilon)\biggr)^2
%+ \psi'(1-\epsilon)-\psi'(2\epsilon)\biggr]
%\]
%\[
\!=\! -2 \psi''(1)+O(\epsilon).
\]
Replacing the above expressions and~(\ref{int_K1}) in Eq.~(\ref{boot1'''}), 
the bootstrap condition takes the form
\[
\left[
\left\{-3 \psi''(1)
+\frac{1}{24} \ln^3\left(\frac{\vec q_1^{\:2}}{\vec q^{\:2}}\right)
- \frac{1}{4}\ln^2\left(\frac{\vec q_1^{\:2}}{\vec q^{\:2}}\right)
\ln\left(\frac{\vec q_1^{\:\prime\:2}}{\vec q^{\:2}}\right)
-\frac{1}{2}I(\vec q_1^{\:\prime\:2},\vec q^{\:2};\vec q_1^{\:2})
\right.\right.
\]
\begin{equation}
\left.\left.
+\frac{1}{2}I(\vec q^{\:2},\vec q_1^{\:2};\vec q_1^{\:\prime\:2})
-\frac{3}{4}J(\vec q_1^{\:2},\vec q^{\:2}; \vec q_1^{\:\prime\:2}) \right\}
+\biggl\{\vec q_1 \leftrightarrow \vec q_1^{\:\prime} \biggr\} \right]
+\frac{1}{\pi}
\int\frac{d\vec q_2}{\vec q_2^{\:2} \vec q_2^{\:\prime\:2}}\:{\cal K}_3 =0\;,
\label{boot1'''''}
\end{equation}
with
\[
I(\vec p_1^{\:2}, \vec p_2^{\:2};(\vec p_1-\vec p_2)^2) =
\frac{(\vec p_1-\vec p_2)^2}{\pi}
\int d\vec p \;\frac{\ln(\vec p_2^{\:2}/\vec p^{\:2})\ln((\vec p_1-\vec p_2)^2
/(\vec p-\vec p_2)^2)}
{(\vec p-\vec p_1)^2(\vec p-\vec p_2)^2}\;,
\]
\[
J(\vec p_1^{\:2}, \vec p_2^{\:2};(\vec p_1-\vec p_2)^2 ) =
\frac{(\vec p_1-\vec p_2)^2}{\pi}
\int d\vec p \;\frac{\ln(\vec p_1^{\:2}/\vec p^{\:2})\ln(\vec p_2^{\:2}/
\vec p^{\:2})}{(\vec p-\vec p_1)^2(\vec p-\vec p_2)^2}\;.
\]
The integrals $I(\vec p_1^{\:2}, \vec p_2^{\:2};(\vec p_1-\vec p_2)^2)$ and
$J(\vec p_1^{\:2}, \vec p_2^{\:2};(\vec p_1-\vec p_2)^2 )$ can be reduced to 
one-dimensional integrals:
\[
I(\vec p_1^{\:2},\vec p_2^{\:2};(\vec p_1-\vec p_2)^2)
=\psi'(1)\ln\vec k_1^{\:2}
+\frac{1}{2}\int_0^1\frac{dx}{x}\ln^2\left(\frac{D_0}{\vec k_2^{\:2}}\right)
\]
\[
+\frac{1}{2}\int_0^1\frac{dx}{1-x}\ln^2\left(\frac{D_0}{\vec k_1^{\:2}}\right)
+\ln \vec k_2^{\:2}\int_0^1\frac{dx}{x}\ln\left(\frac{D_0}{\vec k_2^{\:2}}\right)
+\ln \vec k_1^{\:2}\int_0^1\frac{dx}{1-x}\ln\left(\frac{D_0}{\vec k_1^{\:2}}\right)
\]
\[
+ \int_0^1\frac{dx}{1-x}\ln x \ln D_0
- \int_0^1\frac{dx}{x}\ln x \ln\left(\frac{D_0}{\vec k_2^{\:2}}\right)
- 2 \int_0^1\frac{dx}{1-x}\ln(1-x) \ln\left(\frac{D_0}{\vec k_1^{\:2}}\right)
\]
\begin{equation}
- \int_0^1 dx \ln\left(\frac{x}{1-x}\right)\ln \left(\frac{D_1}{D_0}\right)
\frac{\vec k_1^{\:2}-\vec k_2^{\:2}-(1-2x)}{D_0-D_1}\;,
\label{I}
\end{equation}
\[
J(\vec p_1^{\:2},\vec p_2^{\:2};(\vec p_1-\vec p_2)^2)
=\ln \left(\frac{\vec k_1^{\:2}}{\vec k_2^{\:2}}\right)
\left[\int_0^1\frac{dx}{1-x}\ln\left(\frac{D_0}{\vec k_1^{\:2}}\right)
-\int_0^1\frac{dx}{x}\ln\left(\frac{D_0}{\vec k_2^{\:2}}\right)\right]
\]
\[
+ \int_0^1\frac{dx}{x}\ln^2\left(\frac{D_0}{\vec k_2^{\:2}}\right)
+ \int_0^1\frac{dx}{1-x}\ln^2\left(\frac{D_0}{\vec k_1^{\:2}}\right)
\]
\begin{equation}
- 2\int_0^1 \frac{dx}{D_1-D_0}\ln\left(\frac{x}{1-x}\right)
\ln\left(\frac{D_1}{D_0}\right)
\left[(1-2x)-\frac{D_1(\vec k_1^{\:2}-\vec k_2^{\:2})}{D_0}\right]\;,
\label{J}
\end{equation}
with
\[
\vec k_1 \equiv \frac{\vec p_1}{ |\vec p_1 -\vec p_2|}\;, \;\;\;
\vec k_2 \equiv \frac{\vec p_2}{ |\vec p_1 -\vec p_2|}\;, \;\;\;
D_0\equiv x \vec k_1^{\:2} +(1-x) \vec k_2^{\:2}\;,\;\;\;D_1\equiv x(1-x)\;.
\]
The integral from ${\cal K}_3$ can be re-expressed as follows:  
\begin{equation}
A \equiv \frac{1}{\pi}\int \frac{d^2q_2}{\vec q_2^{\:2} \vec q_2^{\:\prime\:2}}
{\cal K}_3 = (A_1+A_2+A_3)+(\vec q_1 \leftrightarrow \vec q_1^{\:\prime})\;,
\label{A}
\end{equation}
with 
\[
A_1=\frac{1}{2}\int_0^1 \frac{dz}{\vec q^{\:2}(1-z)+ \vec q_1^{\:\prime\:2}z
-\vec q_1^{\:2}z(1-z)}
\]
\[
\times
\left[\left(\frac{2\vec q^{\:2}\vec q_1^{\:2}z}
{\vec q^{\:2}(1-z)+ \vec q_1^{\:\prime\:2}z-\vec q_1^{\:2}z}
+\vec q^{\:2}+\vec q_1^{\:2}- \vec q_1^{\:\prime\:2}\right)\right.
\]
\[
\times\left.
\ln \left(\frac{\vec q^{\:2}(1-z)+ \vec q_1^{\:\prime\:2}z
-\vec q_1^{\:2}z(1-z)}{z(\vec q^{\:2}(1-z)+ \vec q_1^{\:\prime\:2}z)}\right)
\ln \left(
\frac{\vec q^{\:2}(1-z)+ \vec q_1^{\:\prime\:2}z}{\vec q_1^{\:2}z}\right)\right.
\]
\[
\left.+ \left(\frac{2\vec q^{\:2}}{z}-\vec q^{\:2}-\vec q_1^{\:2}
+ \vec q_1^{\:\prime\:2}\right)\ln (1-z)
\ln \left(\frac{\vec q^{\:2}(1-z)+ \vec q_1^{\:\prime\:2}z
-\vec q_1^{\:2}z(1-z)}{\vec q^{\:2}(1-z)}\right)\!\right],
\]
\[
A_2=A_1(\vec q_1^{\:\prime} \leftrightarrow -\vec q)\;,\;\;\;\;\;\;\;\;\;\;
A_3=-A_1(\vec q_1^{\:\prime}=0)\;.
\]
The one-dimensional integrals in Eqs.~(\ref{I}), (\ref{J}) and (\ref{A}) can be 
calculated analytically, but the arising expressions are discouragingly long 
and cumbersome. Nevertheless, the proof can be greatly simplified for the 
following reason. The integrals and the explicit 
logarithms entering the bootstrap condition~(\ref{boot1'''''}) are functions 
of the variables $q_1^2\equiv-\vec q_1^{\:2},\;q_1^{\:\prime\:2} \equiv 
-\vec q_1^{\:\prime\:2}$ and $q^2\equiv -\vec q^{\:2}$. At fixed 
$q_1^{\:2}\leq 0$, $q_1^{\:\prime\:2}\leq 0$, these functions 
are analytical functions of $q^2$, real 
for $q^2 < 0$ and with the cut $0 \leq q^2 < \infty$. Any such function can be 
determined by its discontinuity on the cut, up to another function of $q_1^2$ and 
$q_1^{\:\prime\:2}$ (but not of $q^2$). This last function can be simply 
obtained by evaluating the first function at $\vec q^{\:2}=\infty$.
Operatively, to calculate discontinuities, we have to make the replacement 
$\vec q^{\:2}\rightarrow -q^2-i0$ in the integrals and in the explicit 
logarithms entering the bootstrap condition, to calculate their imaginary 
parts on the upper edge of the cut (they give the discontinuities after 
multiplication by $2i$) and to check the strong bootstrap for the imaginary 
parts (which is the same as for discontinuities). Then, what remains to be
done is to check the bootstrap in the limit $\vec q^{\:2}\gg
\vec q_1^{\:2}, \; \vec q^{\:2}\gg \vec q_1^{\:\prime\:2}$.

The bootstrap relation for imaginary parts (divided by $\pi$) reads
\[
\left\{\left[-\frac{1}{8}\ln^2\left(\frac{\vec q_1^{\:2}}{q^2}\right)
+\frac{5\pi^2}{24}-\frac{1}{2}
\ln\left(\frac{\vec q_1^{\:2}}{q^2}\right)
\ln \left(\frac{\vec q_1^{\:\prime\:2}}{q^2}\right)-\frac{1}{2}\frac{1}{\pi}
\Im I( \vec q_1^{\:2},-q^2-i0; \vec q_1^{\:\prime\:2})
\right.\right.
\]
\[
\left.\left.
+\frac{1}{2}\frac{1}{\pi} \Im I(-q^2-i0,\vec q_1^{\:2};
\vec q_1^{\:\prime\:2} )-\frac{3}{4}\frac{1}{\pi}
\Im J(\vec q_1^{\:2},-q^2-i0; \vec q_1^{\:\prime\:2})\right]+\biggl[\vec q_1^{\:2}
\leftrightarrow \vec q_1^{\:\prime\:2}\biggr]\right\}
\]
\begin{equation}
+\frac{\Im A}{\pi}=0\;.
\label{Im_boot}
\end{equation}
The imaginary part of the integrals $I$ and $J$ in Eqs.~(\ref{I}) and (\ref{J})
can be easily calculated:
\begin{equation}
-\frac{1}{2\pi} \Im I(\vec q_1^{\:2},-q^2-i0; \vec q_1^{\:\prime\:2})
+\frac{1}{2\pi} \Im I(-q^2-i0,\vec q_1^{\:2}; \vec q_1^{\:\prime\:2})
=-\frac{\psi'(1)}{2}+\frac{1}{4}\ln^2\left(\frac{\vec q_1^{\:2}}{ q^2}\right)\!,
\label{ImI}
\end{equation}
\[
\frac{1}{\pi} \Im J(\vec q_1^{\:2},-q^2-i0; \vec q_1^{\:\prime\:2} )+
(\vec q_1^{\:2}\leftrightarrow \vec q_1^{\:\prime\:2})
\]
\begin{equation}
= -\ln\left(\frac{\kappa^-}{q^2}\right)\ln\left(\frac{\kappa^+}{q^2}\right)
+\frac{1}{2}\ln^2\left(\frac{\vec q_1^{\:2}}{q^2}\right)+
\psi'(1)+(\vec q_1^{\:2}\leftrightarrow \vec q_1^{\:\prime\:2})\;.
\label{ImJ}
\end{equation}
To calculate the discontinuity of the integral $A$ defined in Eq.~(\ref{A})
at $q^2=-\vec q^{\:2}\geq 0$, it is convenient to rewrite the integral over $q_2$ in
Minkowski space and to use the Cutkosky rules for the calculation
of the discontinuity. First, we represent the integral 
over $x$ appearing in ${\cal K}_3$ (see Eq.~(\ref{K3}) as
\begin{equation}
I=\int_0^1 dx \int_0^\infty dz \frac{1}{z-k^2x(1-x)-i0}\frac{1}{z-q_1^2(1-x)
-q_2^2x-i0}~,
\end{equation}
where $k, \; q_1, \; q_2$ are considered as vectors in the two-dimensional
Minkowski space, i.e. $k^2=-\vec k^{\:2}, \;q_1^2=-\vec q_1^{\:2},\; q_2^2=
-\vec q_2^{\:2}$. This representation can be used for arbitrary values of
$k^2, \;q_1^2,\; q_2^2$. Analogous representation can be written for
$I(q_i \leftrightarrow
q_i^{\prime})$. It permits to rewrite the integral with ${\cal K}_3$ in the
bootstrap relation in the form
\begin{equation}
A = \frac{1}{i\pi}\int \frac{d^2q_2}{(q_2^{\:2}+i0)((q-q_2)^2+i0)}{\cal K}_3~,
\label{A'}
\end{equation}
where now
\begin{equation}
d^2q_2=dq_2^{(0)}dq_2^{(1)},\;\; q_2^{\:2}=(q_2^{(0)})^2-(q_2^{(1)})^2~,
\end{equation}
etc., which determines $A$ as function of
$q_1^2,\;q_1^{\:\prime\:2}$ and $q^2$ for arbitrary values of
these variables. For $q_1^2\equiv -\vec q_1^{\:2} \leq
0,\;q_1^{\:\prime\:2} \equiv -\vec q_1^{\:\prime\:2} \leq 0$ and
$q^2\equiv -\vec q^{\:2} \leq 0$ it is just the function
entering~(\ref{boot1'''''}), that is easily seen by making the Wick
rotation of the contour of integration over $q_2^{(0)}$. We are
interested in the region $q_1^2\leq 0,\;q_1^{\:\prime\:2} \leq 0$
and $q^2 \geq 0$. According to the Cutkosky rules, the
discontinuity of $A$ related to the terms with $I$ is determined
by the two cuts, with the contributions obtained by the
substitutions:
\begin{equation}
\frac{1}{(q_2^{\:2}+i0)((q-q_2)^2+i0)}\rightarrow (-2\pi i)^2 \delta(q_2^{\:2})
 \delta((q-q_2)^2 )~
\end{equation}
and
\[
\frac{1}{(z-q_1^2(1-x)-q_2^2x-i0)((q-q_2)^2+i0)} 
\]
\begin{equation}
\rightarrow -(-2\pi i)^2
\delta(z-q_1^2(1-x)-q_2^2x)\delta((q-q_2)^2 )~.
\end{equation}
Using these rules and removing the $\delta$-functions by the integration
over $q_2$ (the most appropriate system for this is $q^{(1)}=0,\; q^2=
(q^{(0)})^2$), we obtain
\begin{equation}
\frac{\Im A}{\pi}=-\frac{3}{2}\ln\left(\frac{\kappa^-}{q^2}\right)
\ln\left(\frac{\kappa^+}{q^2}\right)
+\frac{1}{4}\ln^2\left(\frac{\vec q_1^{\:2}\vec q_1^{\:\prime\:2}}{(q^2)^2}
\right)+\frac{1}{2}
\ln\left(\frac{\vec q_1^{\:2}}{q^2}\right)
\ln\left(\frac{\vec q_1^{\:\prime\:2}}{q^2}\right)\;,
\label{ImA}
\end{equation}
with
\[
\kappa^{\pm}=\frac{1}{2}\left(q^2+\vec q_1^{\:2}+\vec q_1^{\:\prime\:2}
\pm \sqrt{(q^2+\vec q_1^{\:2}+\vec q_1^{\:\prime\:2})^2-4\vec q_1^{\:2}
\vec q_1^{\:\prime\:2}}\right)\;.
\]
Using Eqs.~(\ref{ImI}), (\ref{ImJ}) and (\ref{ImA}), it is easy to see that 
the imaginary part of the bootstrap relation, Eq.~(\ref{Im_boot}), is satisfied. 
Finally, the bootstrap has to be considered in the limit $\vec q^{\:2}\gg 
\vec q_1^{\:2}, \; \vec q^{\:2}\gg \vec q_1^{\:\prime\:2}$. 
We have in this limit
\[
I( \vec q_1^{\:2},\vec q^{\:2}; \vec q_1^{\:\prime\:2})\simeq -\zeta(2)
\ln\left(\frac{\vec q^{\:2}}{\vec q_1^{\:2} }\right)+2\zeta(3)\;,
\]
\[
I( \vec q^{\:2},\vec q_1^{\:2}; \vec q_1^{\:\prime\:2})\simeq -\frac{1}{6}
\ln^3\left(\frac{\vec q^{\:2}}{\vec q_1^{\:2} }\right)-\zeta(2)
\ln\left(\frac{\vec q^{\:2}}{\vec q_1^{\:2} }\right)+2\zeta(3)\;,
\]
\[
J( \vec q_1^{\:2},\vec q^{\:2}; \vec q_1^{\:\prime\:2})\simeq -\frac{1}{6}
\ln^3\left(\frac{\vec q^{\:2}}{\vec q_1^{\:2} }\right)-2\zeta(2)
\ln\left(\frac{\vec q^{\:2}}{\vec q_1^{\:2} }\right)+4\zeta(3)
\]
and
\[
\frac{1}{\pi}
\int\frac{d\vec q_2}{\vec q_2^{\:2} \vec q_2^{\:\prime\:2}}\:{\cal K}_3\simeq
-\frac{1}{4} \ln \left(\frac{\vec q^{\:2}}{\vec q_1^{\:\prime\:2} }\right)
\ln\left(\frac{\vec q^{\:2}}{\vec q_1^{\:2} }\right)\left( \ln
\left(\frac{\vec q^{\:2}}{\vec q_1^{\:2} }\right)+\ln \left(\frac{\vec q^{\:2}}
{\vec q_1^{\:\prime\:2} }\right) \right)
\]
\[
-\frac{3\zeta(2)}{2}\left(
\ln \left(\frac{\vec q^{\:2}}{\vec q_1^{\:2} }\right)+
\ln \left(\frac{\vec q^{\:2}}{\vec q_1^{\:\prime\:2} }\right)\right)-6 \zeta(3)\;.
\]
Again, it is easy to see that the bootstrap condition~(\ref{boot1'''''})
is satisfied in the limit of large $\vec q^{\:2}$.

%\section{Conclusions}
%
%The basic assumption of the BFKL theory is gluon Reggeization. This 
%property has been rigorously proved so far only in the LLA.
%In the NLA it is only a hypothesis which needs to be verified.
%The self-consistency of the Reggeized form of elastic amplitudes leads
%to soft bootstrap conditions. Their fulfillment, already
%proved, is a stringent test of the hypothesis of gluon Reggeization in the NLA.
%
%Bootstrap conditions must be satisfied also for the inelastic amplitudes
%involved in the derivation of the BFKL equation. Among these conditions,
%there are the ``strong'' bootstrap conditions just proved in the NLA.
%Their fulfillment supports in an even stricter way the hypothesis of gluon 
%Reggeization in the NLA and opens the way to a rigorous proof of this property
%in the NLA.


\begin{thebibliography}{99}

\bibitem{Fad02}
Fadin, V.S (2002) these proceedings and references therein.

\bibitem{FF98}
Fadin, V.S. and Fiore, R. (1998) The generalized nonforward BFKL equation and the
'bootstrap' condition for the gluon Reggeization in the NLLA, 
{\it Phys. Lett.}, {\bf Vol.~no.~B440}, pp.~359--366

\bibitem{FFP99}
Fadin, V.S., Fiore, R. and Papa, A. (1999) The quark part of the nonforward BFKL 
kernel and the 'bootstrap' for the gluon Reggeization, {\it Phys. Rev.},
{\bf Vol.~no.~D60}, 074025 (13 pages)

\bibitem{FFK00}
Fadin, V.S., Fiore, R. and Kotsky, M.I. (2000) The compatibility of the 
gluon Reggeization with the $s$ channel unitarity, {\it Phys. Lett.}, 
{\bf Vol.~no.~B494}, pp.~100--108

\bibitem{Kot02}
Kotsky, M.I. (2002) these proceedings

\bibitem{FFKP00_Q}
Fadin, V.S., Fiore, R., Kotsky, M.I. and Papa, A. (2000) Quark impact factors,
{\it Phys. Rev.}, {\bf Vol.~no.~D61}, 094006 (16 pages)

\bibitem{FFKP00_G}
Fadin, V.S., Fiore, R., Kotsky, M.I. and Papa, A. (2000) Gluon impact factors,
{\it Phys. Rev.}, {\bf Vol.~no.~D61}, 094005 (22 pages)

\bibitem{BV99} 
Braun, M. (1999) Comments on the second order bootstrap relation, 
hep-ph/9901447; Braun, M. and Vacca, G.P. (2000) The bootstrap
for impact factors and the gluon wave function, {\it Phys. Lett.}, 
{\bf Vol.~no.~B447}, pp.~156--162

\bibitem{FFKP00}
Fadin, V.S., Fiore, R., Kotsky, M.I. and Papa, A. (2000) Strong bootstrap 
conditions, {\it Phys. Lett.}, {\bf Vol.~no.~B495}, pp.~329--337

\bibitem{FP02}
Fadin, V.S. and Papa, A. (2002) A proof of fulfillment of the strong bootstrap 
condition, {\it Nucl. Phys.}, {\bf Vol.~no.~B640}, pp.~309--330

\end{thebibliography}
\end{document}